\documentclass{article}

\usepackage{amsmath}
\usepackage{amssymb}
\usepackage{graphicx}

\newcommand{\coll}[1]{\ensuremath{\left\{ {#1}\right\} }}

\newcommand{\ket}{\rangle}
\newcommand{\sem}[1]{\ensuremath{[\![{#1}]\!]}}

\newcommand{\paren}[1]{\ensuremath{\left( {#1} \right)}}
\newcommand{\brac}[1]{\ensuremath{\left[ {#1} \right]}}

\newcommand{\set}[2]{\ensuremath{\left\{\left.#1\,\,\vphantom{#2}\right|\,#2\right\}}}
\newcommand{\fall}[1]{{\forall\,{#1},\ }}

\newcommand{\mc}[1]{{\mathcal{#1}}}

\newcommand{\mb}[1]{{\bf #1}}

\newcommand{\dashVdash}{\dashv\vdash}

\begin{document}



\title{Dynamic Quantum Logic for Quantum Programs}

\author{Olivier Brunet, Philippe Jorrand \\ Leibniz Laboratory, University of Grenoble \\ 46, avenue F\'elix Viallet, 38000 Grenoble, France \\ \tt olivier.brunet@imag.fr \\ \tt philippe.jorrand@imag.fr}

\maketitle


\begin{abstract}
We present a way to apply quantum logic to the study of quantum programs. This is made possible by using an extension of the usual propositional language in order to make transformations performed on the system appear explicitly. This way, the evolution of the system becomes part of the logical study. We show how both unitary operations and two-valued measurements can be included in this formalism and can thus be handled logically.
\end{abstract}


\section{Introduction}

The logical study of quantum mechanics, originated in the thirties by von Neumann and Birkhoff\cite{Birkhoff36QuantumLogic}, aims at investigating formally what makes quantum mechanics so different from the classical world. To quote the pioneering article:
\emph{``{\it One of the aspects of quantum theory which has attracted the most general attention, is the novelty of the logical notions which it presupposes... The object of the present paper is to discover what logical structures one may hope to find in physical theories which, like quantum mechanics, do not conform to classical logic.}''}
The starting point of this study is based on the use of {\em closed subspaces} of a Hilbert space $\mc H$ for representing properties about the system. The operations defined on subspaces, such as the orthocomplementation and the intersection, are interpretations of the negation and conjunction on propositions, thus allowing to define a full-fledged propositional logic. This constitutes the {\em standard quantum logic} or {\em orthomodular quantum logic}\cite{Hughes89Book,Svozil98Book,DallaChiara2001QuantumLogic}.

Since its origins, many variations have been studied, and different attempts have been made to identify some axioms or conditions which would permit to recapture the Hilbert space formalism\cite{Mackey57Book,Piron76Book,Ptak91Book}. Unfortunately, despite the large amount of publications on this topic, these works have remained extremely theoretical, and have led to very little applications. However, it is possible to use the quantum logic formalism to express and study properties in a quantum computation context, by extending the language in order to have quantum operations appear explicitly and thus having the possibility to include the evolution of a system in the logical study. 

In the present article, we present such a kind of extension of the quantum logic formalism. It is based on the use of closed subspaces as partial descriptions of states of the system (with statements of the form ``the actual state lies in this subspace''), and logical assertions can then be seen as relating knowledge about the system's state at different moments during the computation. With this approach, we will consider the application of unitary operators, which modify the knowledge without losing information, and the process of measurement, where some loss of information occurs.

\section{Standard Quantum Logic}

\subsection{Basic definitions}

Our formalism for representing subspaces by terms of a propositional language relies on two elements: a language (i.e{.} a set of propositional terms) $\mc L$ and an {\em interpretation} function $\sem \cdot$ which maps each term of the language to a closed subspace of $\mc H$. This way, each term $p \in \mc L$ is associated to a closed subspace of $\mc H$, denoted $\sem p$. Our language $\mc L$ contains two connectives: negation $\neg$ and conjunction $\wedge$, and a set of constants $\Psi$. Thus, every constant $p \in \Psi$ is a term (i.e. $p \in \mc L$) and given two terms $p,q \in \mc L$, both $\neg p$ and $p \wedge q$ are terms.

The definition of the interpretation function is based on the structure of the terms, and on the correspondance between the negation $\neg \cdot$ and the orthocomplementation $\cdot^\bot$ on the one hand, and between the conjunction $\cdot \wedge \cdot $ and the intersection $\cdot \cap \cdot $ on the other hand. Thus, the definition of $\sem \cdot$ is given by:
\begin{equation}
\fall {p \in \mc L} \sem {\neg p} = \sem p^\bot \hskip1cm \fall {p,q \in \mc L} \sem {p \wedge q} = \sem p \cap \sem q
\end{equation}
The definition is completed by the interpretation of each atomic proposition. In the following, we will consider the restricted case where the Hilbert space $\mc H$ is of the form $\otimes^n \mb C^2$, and atomic propositions are $z_i$ and $x_i$ with $1 \leq i \leq n$. Intuitively, the propositions relate to the corresponding direction of the $i^{th}$ qubit. If $n=1$, the interpretations are defined by $\sem z = \mb C |1\ket$ and $\sem x = \mb C |-\ket = \mb C \paren{|0\ket - |1\ket}$, and for $n>1$, this definition is extended by using tensor products. For instance: 
 \begin{equation}
 \sem {x_i} = \paren {\otimes^{i-1} \mb C^2} \otimes \mb C |-\ket \otimes \paren {\otimes^{n-i} \mb C^2}
 \end{equation}
Finally, two constants are useful to define: the true proposition $\top$, verified everywhere (its interpretation $\sem \top$ equals the whole Hilbert space $\mc H$) and the absurd proposition $\bot$ which cannot be verified, so that $\sem \bot = \coll 0$.

\subsection{Additional connectives}

Even though the logical language as defined above is expressive enough, it is interesting to introduce other connectives, using the two already available operations (negation $\neg$ and conjunction $\wedge$). First, we define the disjunction $p \vee q$ as $\neg (\neg p \wedge \neg q)$. In terms of subspaces, this connective has a simple formulation, since it corresponds to the sum of two subspaces:
\begin{equation}
\sem {p \vee q} = \sem {\neg \paren{\neg p \wedge \neg q}} = \paren{\sem p^\bot \cap \sem q^\bot}^\bot = \sem p \oplus \sem q
\end{equation}
An implication connective $ p \rightarrow q $ can also be defined, using its classical definition, that is $ \neg p \vee q $ or equivalently $ \neg ( p \wedge \neg q)$, so that $\sem {p \rightarrow q} = \sem p^\bot \oplus \sem q$. With this connective, it is easy to define the equivalence connective: $p \leftrightarrow q$ stands for $(p \rightarrow q) \wedge (q \rightarrow p)$. This connective will be very useful in our approach, as it can be used to express some kind of equality between different qubit states. Finally, we also introduce the exclusive disjunction connective $p \veebar q$ as the negation of the equivalence, that is $\neg (p \leftrightarrow q)$. Its interpretation can be expressed as:
\begin{equation}
\sem {p \veebar q} = \paren{\sem p \wedge \sem q^\bot} \oplus \paren{\sem p^\bot \wedge \sem q}
\end{equation}
This operator appears frequently in the study of quantum program, since it is the logical equivalent to the addition modulo 2 for integers.

\subsection{Example: Description of an e.p.r. pair} \label{Ex:EPRDesc}

This simple logical language permits to fully describe many interesting states of a quantum system. To illustrate this, we show that proposition $(z_1 \leftrightarrow z_2) \wedge (x_1 \leftrightarrow x_2)$ is a complete description of an {\em e.p.r.} pair\cite{Einstein35EPR,NielsenChung2000Book}:
\begin{equation*}
\begin{array}{rl}
\sem {z_1 \leftrightarrow z_2}
& = \sem {z_1\rightarrow z_2} \wedge \sem {z_2 \rightarrow z_1} \\
& = \paren { \sem {z_1}^\bot \oplus \sem {z_2} } \cap \paren {\sem {z_1} \oplus \sem {z_2}^\bot } \\
& = \paren {\mb C |00\ket \oplus \mb C |10\ket \oplus \mb C |11\ket} \oplus \paren{\mb C |00\ket \oplus \mb C |01\ket \oplus \mb C |11\ket} \\
& = \mb C |00\ket \oplus \mb C |11\ket \\
& \ \\
\sem {x_1 \leftrightarrow x_2} & = \mb C |{+}+\ket \oplus \mb C |{-}-\ket \\
& \ \\
\sem {(z_1 \leftrightarrow z_2) \wedge (x_1 \leftrightarrow x_2)}
& = \sem {z_1 \leftrightarrow z_2} \cap \sem {x_1 \leftrightarrow x_2} \\
& = (\mb C |00\ket \oplus \mb C |11\ket) \cap (\mb C |{+}+\ket \oplus \mb C |{-}-\ket) \\
& = \mb C (|00\ket + |11\ket)
\end{array}
\end{equation*}

Equivalently, one can use $\paren{z_1 \veebar z_2 \leftrightarrow \bot} \wedge \paren{x_1 \veebar x_2 \leftrightarrow \bot}$ to describe e.p.r. pairs. In that case, one can interpret $\veebar$ as the addition modulo 2, $\leftrightarrow$ as the equality and $\bot$ as $0$. Similarly, it can be shown that proposition $(z_1 \veebar z_2 \leftrightarrow \bot) \wedge (z_1 \veebar z_3 \leftrightarrow \bot) \wedge (x_1 \veebar x_2 \veebar x_3 \leftrightarrow \bot)$ is a complete characterization of a GHZ state.

\subsection{Entailment} 

In order to be able to relate propositions seen as different descriptions of a system, we introduce a last notion, corresponding to the inclusion of interpretations: given two terms $p$ and $q$, $p$ will be said to entail $q$ (which we will denote $ p \Vdash q$) if and only if $ \sem p \subseteq \sem q$. If both interpretations are equal, we may also write $ p \dashVdash q$.

This entailment relation can be related to the implication connective, and more precisely, to terms $p$ and $q$ verifying $\sem {p \rightarrow q} = \mc H$, which can be written as $ \top \Vdash p \rightarrow q$, or more shortly $\Vdash p \rightarrow q$.  For instance, it can be easily shown that if $p \Vdash q$, then $\Vdash p \rightarrow q$, but the converse is not true, as illustrated by the fact that for a single qubit, one has $\Vdash z \rightarrow x$ but $\sem z \not\subseteq \sem x$.

Contrary to the implication, it is possible to express and perform deductions using the entailment relation: from $p \Vdash q$ and $q \Vdash r$, it is possible to deduce that $p \Vdash r$. This motivates the fact that this relation will be used in the following to express relations among properties verified by a system at different steps of a quantum program.

\section{Dynamic Aspects, Unitary Operations}

\subsection{Extension of the language}

In order to include an explicit reference to the dynamic evolution of a system, we will extend our propositional language by adding a collection of unary connectives (denoted $\brac u$), each corresponding to the application of an unitary operator $U$ on the system. The idea is to associate a proposition $\brac u p$ to a system initially verifying $p$ (that is in a state $|\varphi\ket$ in $\sem p$) and on which $U$ is applied. This permits to define the interpretation of such a connective:
\begin{equation}
\sem {\brac u p} = \set {U |\varphi\ket}{|\varphi\ket \in \sem p}
\end{equation}

With the introduction of these additional connectives, it becomes possible to express relationships between the different states of a system along the execution of unitary transformations. For instance, simple calculations show that the subspace spanned by $|1\ket$ is left unchanged by the application of the Hadamard operator, since $\sigma_z |1\ket = -|1\ket$. This can be written logically as: $ \brac {\sigma_z} z \dashVdash z$. Similarly, one has $ \brac {\sigma_y} x \dashVdash \neg x$, since $\sigma_y |-\ket = i |+\ket$. It is possible to express similar assertions for more complex propositions. For instance:
\begin{equation}
\brac{\oplus_{1,2}} (z_2 \leftrightarrow \bot) \dashVdash z_1 \leftrightarrow z_2
\end{equation}
The linearity and invertibility of unitary operators implies that the application of such an operator does commute with both orthocomplementation and intersection operations. Logically, one can thus write:
$$ \brac u (\neg p) \dashVdash \neg (\brac u p) \hskip2cm \brac u (p\wedge q) \dashVdash (\brac u p) \wedge (\brac u q) $$
This means in particular that the definition of the behaviour of different operators can be done by just specifying their behaviour for atomic propositions. For instance, the complete description of $\sigma_{z,i}$ (where the $i$ indice means that $\sigma_z$ acts on the $i$th qubit) for atomic propositions $z$ and $x$ is given by:
\begin{equation}
\brac {\sigma_{z,i}} z_i \dashVdash z_i \hskip1cm \brac {\sigma_{z,i}} x_i \dashVdash \neg x_i
\end{equation}
Properties corresponding to other qubits are left unchanged. For instance, one has: $\brac {\sigma_{z,1}} x_2 \dashVdash x_2 $. With atomic terms $z$, $x$ and $y$ for each qubit, it is possible to provide the complete description of many common operators, such as the Pauli and Hadamard operators, the controlled-not and the Toffoli operator\cite{barenco95elementary,DiVincenzo98Circuits}. They are given in figure 1.

\begin{figure}
\hrule
\vspace{5pt}
Pauli operators
\begin{equation*}
\begin{array}{rl@{\hskip1cm}rl}
\brac {\sigma_{z,i}} z_i & \dashVdash z_i & \brac {\sigma_{z,i}} x_i & \dashVdash \neg x_i \\
\brac {\sigma_{x,i}} z_i & \dashVdash \neg z_i & \brac {\sigma_{x,i}} x_i & \dashVdash x_i \\
\brac {\sigma_{y,i}} z_i & \dashVdash \neg z_i & \brac {\sigma_{y,i}} x_i & \dashVdash \neg x_i
\end{array}
\end{equation*}
Hadamard operator
\begin{equation*}
\begin{array}{rl@{\hskip1cm}rl}
\brac {H_i} z_i & \dashVdash x_i & \brac {H_i} x_i & \dashVdash z_i
\end{array}
\end{equation*}
Controlled-Not operator
\begin{equation*}
\begin{array}{rl@{\hskip1cm}rl}
\brac {\oplus_{i,j}} z_i & \dashVdash z_i & \brac {\oplus_{i,j}} x_i & \dashVdash x_i \veebar x_j \\
\brac {\oplus_{i,j}} x_j & \dashVdash x_j & \brac {\oplus_{i,j}} z_j & \dashVdash z_j \veebar z_i
\end{array}
\end{equation*}
Toffoli operator
\begin{equation*}
\begin{array}{rl@{\hskip1cm}rl}
\brac {T_{i,j,k}} z_i & \dashVdash z_i & \brac {T_{i,j,k}} x_i & \dashVdash x_i \veebar (z_j \wedge x_k) \\
\brac {T_{i,j,k}} z_j & \dashVdash z_j & \brac {T_{i,j,k}} x_j & \dashVdash x_i  \veebar (z_i \wedge x_k) \\
\brac {T_{i,j,k}} x_k & \dashVdash x_k & \brac {T_{i,j,k}} z_k & \dashVdash z_k  \veebar (z_i \wedge z_j)
\end{array}
\end{equation*}
\vspace{5pt}
\hrule
\label{Tab:Unitary} \caption{Definition of some usual unitary operators}
\end{figure}

It should be noted that since the action of the Toffoli operator can be described, it follows that this formalism is more general that than of stabilizers which plays a central role in the Gottesman-Knill theorem\cite{NielsenChung2000Book}, if one considers unitary operations only. However, we will see in section \ref{Sec:Measure} how measurements can be included in our formalism.

\subsection{Example: Creation of an {\em epr} pair} \label{Ex:EPRCreat}

\begin{figure} \label{Fig:EPR}
\hrule
\vspace{5pt}
\center{\includegraphics[width=4cm]{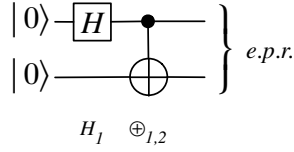}}
\vspace{5pt}
\hrule
\caption{e.p.r. pair creation circuit}
\end{figure}

The usual process for creating an {\em epr} pair is to start from $|00\ket$ (which is logically expressed as $\neg z_1 \wedge \neg z_2$ or equivalently as $(z_1 \leftrightarrow \bot) \wedge (z_2 \leftrightarrow \bot)$) to apply $H_1$ and then $\oplus_{1,2}$ to the system, as represented in figure 2. Logically, the quantum circuit can be studied by the following calculation:

\begin{equation*}
\begin{array}{rl}
\brac {H_1} ( \neg z_1 \wedge \neg z_2 )
& \dashVdash \brac {H_1} ( \neg z_1 ) \wedge \brac {H_1} ( \neg z_2 ) \\
& \dashVdash \neg (\brac {H_1}  z_1 ) \wedge \neg (\brac {H_1} z_2) \\
& \dashVdash \neg x_1 \wedge \neg z_2 \\
\end{array}
\end{equation*}
\begin{equation*}
\begin{array}{rl}
\brac{\oplus_{1,2}} \brac {H_1} ( \neg z_1 \wedge \neg z_2 )
& \dashVdash \brac{\oplus_{1,2}} ( \neg x_1 \wedge \neg z_2 ) \\
& \dashVdash \neg (\brac{\oplus_{1,2}} x_1) \wedge \neg (\brac{\oplus_{1,2}} z_2 ) \\
& \dashVdash \neg (x_1 \veebar x_2) \wedge \neg (z_1\veebar z_2 ) \\
& \dashVdash \paren{x_1 \leftrightarrow x_2} \wedge \paren{z_1\leftrightarrow z_2}
\end{array}
\end{equation*}

As expected, the final proposition, that is $\paren{x_1 \leftrightarrow x_2} \wedge \paren{z_1\leftrightarrow z_2}$, provides a complete characterization of the subspace spanned by e.p.r. pairs as we have seen in example \ref{Ex:EPRDesc}.

\subsection{Example: A teleportation circuit}

In order to illustrate the use of our formulation of quantum logic for the study of more complex  programs, we develop a teleportation circuit\cite{Bennett93Teleportation,brassard98teleportation} and show how it is possible to relate properties verified by a qubit before and after the teleportation. The circuit is defined in figure 3.
\begin{figure} \label{Fig:Teleportation}
\hrule
\vspace{5pt}
\center{\includegraphics[width=10.4cm]{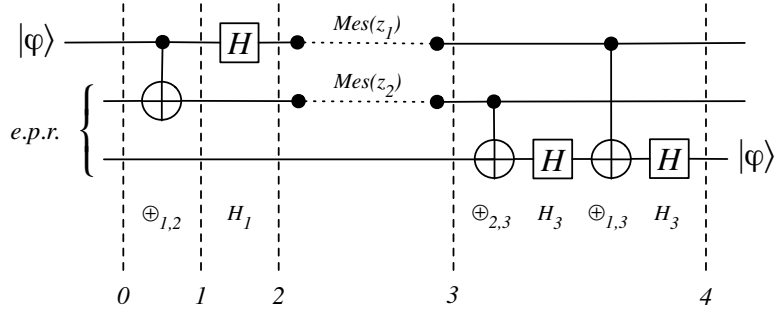}}
\vspace{5pt}
\hrule
\caption{Teleportation circuit}
\end{figure}

Let us first concentrate on the left part of the circuit, from step $0$ to step $2$. One has for the first qubit:
\begin{equation}
z_1 \dashVdash \brac {H_1} x_1 \dashVdash \brac{\oplus_{1,2}} \brac{H_1} \paren{x_1 \veebar x_2}
\end{equation}
We simplify these notations by using exponents to indicate the stage of the computation (i.e. the number of the vertical dashed lines). This permits to remove the unary connectives corresponding to the gates, so that for instance the previous proposition rewrites as:
\begin{equation}
z_1^2 \dashVdash x_1^1 \dashVdash x_1^0 \veebar x_2^0
\end{equation}
Likewise, the second qubit verifies $z_2^2 \dashVdash z_1^0 \veebar z_2^0 $. The third qubit is left unchanged, so that $z_3^2 \dashVdash z_3^0 $ and $x_3^2 \dashVdash x_3^0$. Now, since qubits $2$ and $3$ were part in the beginning of an e.p.r. pair, it follows that $z_2^0 \dashVdash z_3^0$ and $x_2^0 \dashVdash x_3^0$. From this, some manipulations can be done on properties, so that the first part of the system can be fruitfully characterized by these two propositions:
\begin{equation}
x_1^0 \dashVdash x_3^2 \veebar z_1^2 \hskip1cm z_1^0 \dashVdash z_3^2 \veebar z_2^2
\end{equation}

Concerning the second part of the circuit, similar calculations permit to express the following properties about the third qubit :
\begin{equation}
z_3^4 \dashVdash z_2^3 \veebar z_3^3 \hskip1cm x_3^4 \dashVdash z_1^3 \veebar x_3^3
\end{equation}

If we remove the measurements and bit-transmissions and identify steps 2 and 3, these two portions can be combined, and lead to:
\begin{equation}
z_3^4 \dashVdash z_1^0 \hskip1cm x_3^4 \dashVdash x_1^0
\end{equation}
This shows that properties $z$ and $x$ on the first qubit at the beginning are transformed into the same properties on the third qubit at the end of the circuit.

\section{Dealing with Measurements} \label{Sec:Measure}

\subsection{A new unary connective}

We have explained how unitary operators can be included in our logical formalism by the introduction of unary connectives and explore now the way measurements can be expressed in our formalism. For simplicity, we will consider only one form of measurement, that of qubit $i$ along the $z$-direction, which we will represent by a collection of unary connectives $\brac{m_z(i)}$.

One needs, given a proposition $p$, to determine the interpretation of $\brac{m_z(i)}p$. This is done as before, with $\sem{\brac{m_z(i)}p}$ defined as the smallest subspace containing every state which can be obtained after the measurement when starting from elements of $\sem p$.

For this, suppose that our system is in a state $|\varphi\ket$ and verifies property $p$ (so that $|\varphi\ket \in \sem p$) and let us perform the measurement of the first qubit along $z$. After that, the state $|\varphi\ket$ has been transformed either into state $\frac 1 2 (|\varphi\ket + \sigma_z |\varphi\ket)$ or $\frac 1 2 (|\varphi\ket - \sigma_z |\varphi\ket)$ (rigourously, if the system is made of $n$ qubits, one should write $\sigma_z \otimes I^{n-1}$ but we voluntarily use a simplified notation, since it does not add any ambiguity).
Thus, if proposition $p$ represents knowledge about the system before the measurement, the new state belongs to the set:
\begin{equation}
S_p = \set {\frac 1 2 (|\varphi\ket+\sigma_z|\varphi\ket)}{|\varphi\ket \in \sem p} \cup \set {\frac12(|\varphi\ket-\sigma_z|\varphi\ket)}{|\varphi\ket \in \sem p}
\end{equation}
But propositions are represented by closed subspaces, so that $\sem{\brac{m_z(1)}p}$ is actually the subspace spanned by $S_p$. Now, let $|\varphi\ket$ be a state in $\sem p$. From its definition, $S_p$ contains both $\frac 1 2 (|\varphi\ket + m|\varphi\ket)$ and $\frac 1 2 (|\varphi\ket - m|\varphi\ket)$, so that by additivity $|\varphi\ket \in \sem{\brac{m_z(1)}p}$. It follows that $\sem p \subseteq {\rm span}(S_p)$. Similarly, considering the difference, $\sigma_z |\varphi\ket$ is also in $\sem{\brac{m_z(1)}p}$. We have thus shown that:
\begin{equation}
\sem {p \vee \brac {\sigma_{z,1}} p} = \sem p \oplus \sem {\brac {\sigma_{z,1}} p} \subseteq \sem{\brac{m_z(1)}p}
\end{equation}
Conversely, if $|\varphi\ket$ is in $\sem p$, then $\frac 1 2 (|\varphi\ket \pm \sigma_z |\varphi\ket) \in \sem {p \vee \brac {\sigma_{z,1}} p}$, which implies by linearity that actually, one has:
\begin{equation}
\sem {p \vee \brac {\sigma_{z,1}} p} = {\rm span}(S_{p}) = \sem{\brac{m_z(1)}p}
\end{equation}

Thus, we have shown that starting from a system verifying property $p$ and after measuring its first qubit along $z$, the most precise proposition describing the system is $p \vee \brac {\sigma_{z,1}} p$. This result can be generalized to other qubits, and one has: 
\begin{equation}
\brac{{m_z}(i)} p \dashVdash p \vee \brac {\sigma_{z,i}} p
\end{equation}
It should be noted that the measurement need not be restricted to the $z$-direction of a qubit. Actually, any hermitian operator $o$ which has $\pm 1$ as eigenvalues and is formalizable in our logic can be used to define a measurement operation, which interpretation for a proposition $p$ would then be equivalent to that of $p \vee \brac o p$.

\subsection{Example: A teleportation circuit, continued}

Now that we have introduced measurements in our formalism, we can finish the study of the previous example, by expressing the relations between properties at points $2$ and $3$. Consider for instance the way proposition $x_3 \veebar z_1$ is transformed during a measurement of qubit $1$ along $z$:
\begin{equation}
\begin{array}{r@{}l}
\brac{m_z(1)}\paren{x_3 \veebar z_1}
& \dashVdash \paren{x_3 \veebar z_1} \vee \brac{\sigma_{z,1}}\paren{x_3 \veebar z_1} \\
& \dashVdash \paren{x_3 \veebar z_1} \vee \paren{\brac{\sigma_{z,1}} x_3 \veebar \brac{\sigma_{z,1}} z_1} \\
& \dashVdash \paren{x_3 \veebar z_1} \vee \paren{x_3 \veebar z_1} \\
& \dashVdash \paren{x_3 \veebar z_1}
\end{array}
\end{equation}
From this, we deduce $x_3^2 \veebar z_1^2 \dashVdash x_3^3 \veebar z_1^3$, and similarly, one has $z_3^2 \veebar z_2^2 \dashVdash z_3^3 \veebar z_2^3$, so that the measurement process does not affect our program in the sense that regarding properties, their succession is the same as if one had a simple wire instead. As a consequence, the expected relations between the first qubit at step $0$ and the third qubit at step $4$ still hold despite the measurement:
\begin{equation}
z_3^4 \dashVdash z_1^0 \hskip1cm x_3^4 \dashVdash x_1^0
\end{equation}

\subsection{Measurements and partial representations}

Thanks to its simple logical characterization, it is possible to express interesting properties about connective $\brac{m_z(i)}$. A first remark that can be done is that performing a measurement on the system acts for propositions as an approximation operation, so that the result is less informative than the starting argument. In other words, the interpretation of the result contains the interpretation of the initial proposition:
\begin{equation}
p \Vdash \brac{m_z(i)} p {\rm \ \ \ or \ equivalently\ \ \ } \sem p \subseteq \sem{ \brac{m_z(i)} p }
\end{equation}
In some situations, no information is lost (for instance, $ \brac{m_z(1)} z_1 \dashVdash z_1$) whereas it might also happen that every information is lost, leading to $\top$ as a result: $ \brac{m_z(1)} x_1 \dashVdash x_1 \vee \neg x_1 \dashVdash \top $. This illustrates the irreversibility of the measurement process.

Moreover, relation $\Vdash$ can be seen as a partial order, making operation $\brac{m_z(i)}$ monotonous and idempotent, that is $\brac{m_z(i)} \brac{m_z(i)} p$ and $\brac{m_z(i)} p$ are equivalent with regards to $\dashVdash$. These three properties form the definition of {\em upper closure operators}, which are a general formalization of the notion of approximation.
This suggests to envision propositions about the system as partial descriptions of its state. From this point of view, measurements correspond to loss of information and unitary operation to transformation of information (with neither loss nor addition).

Addition of information can also be formalized using conjunctions. This situation arises for instance after measurements, when one takes into account the result of the measurement. Starting from a proposition $p$, the resulting proposition then becomes either $\brac{m_z(i)}p \wedge z_i$ or $\brac{m_z(i)}p \wedge \neg z_i$. And since a system cannot verify the absurd proposition $\bot$, this type of construction provides some informations about the possibility of a given outcome, since for instance if $\brac{m_z(i)}p \wedge z_i \dashVdash \bot$, then the outcome corresponding to $z_i$ can not occur. A important example for this is when starting from a proposition $p$ such that $p \Vdash z_i$ (which can be interpreted as {\em ``one knows that $z_i$ holds''}), one has $\brac{m_z(i)} p \Vdash z_i$ (by monotony of $\brac{m_z(i)}$ and the fact that $z_i \dashVdash \brac{m_z(i)} z_i$) so that $\brac{m_z(i)} p \wedge \neg z_i \Vdash \bot$, meaning that outcome $\neg z_i$ is not possible.

This discussion shows that it is rather natural to consider knowledge about a quantum system from a partial description point of view, and that it is possible to describe the behaviour of usual operations in terms of knowledge. 

\section{Conclusion}

In this article, we have shown how the basic quantum logic formalism can be extended into a dynamic quantum logic by the addition of several unary connectives which do all correspond to an action that can be performed of a quantum system. This provides a method for the logical study of quantum programs.

A few comments can be done about this approach. First, it is purely non-statistical, so that for measurements in particular, no information is provided about the probability of a particular outcome. This problem could be studied by, for instance, adding probability measures on the different subspaces or equivalently on properties.

Moreover, this approach suffers from the use of orthomodular quantum logic as underlying logic. This logic is extremely uneasy to manipulate, due to the fact that the distributivity of disjunction over conjunction and vice versa do not hold. A weaker property, called orthomodularity, holds but does not permit efficient formula manipulations. As a result, during a computation, the size of propositions tend to grow exponentially. A solution to this problem is suggested by the fact that, as developed in section 4.3, a interesting approach is to view properties as partial descriptions. In that case, a convenient logic is provided by intuitionnistic logic, a non-classical logic which main specificity is that the excluded middle principle ($\varphi \vee \neg \varphi$) does not hold. The advantage would be the obtention of a distributive and decidable logic for representing and studying quantum programs. This is the type of approach that we have started to investigate in Ref.~\cite{Brunet03Completion}.

\end{document}